# Network Properties for Robust Multilayer Infrastructure Systems: A Percolation Theory Review

ZAHRA MAHABADI, LIZ VARGA, AND TOM DOLAN
Department of Civil, Environmental and Geomatic Engineering, University College London, London WC1E 6BT, U.K.
Corresponding authors: Zahra Mahabadi (zahra.mahabadi.19@ucl.ac.uk) and Liz Varga (l.varga@ucl.ac.uk)

The work of Zahra Mahabadi was supported by the Department of Civil, Environmental and Geomatic Engineering, University College London. The work of Liz Varga and Tom Dolan was supported by the Engineering and Physical Sciences Research Council (EPSRC) under Grant EP/R017727/1 relating to the U.K. Collaboratorium for Research on Infrastructure and Cities (UKCRIC).

**ABSTRACT** Infrastructure systems, such as power, transportation, telecommunication, and water systems, are composed of multiple components which are interconnected and interdependent to produce and distribute essential goods and services. So, the robustness of infrastructure systems to resist disturbances is crucial for the durable performance of modern societies. Multilayer networks have been used to model the multiplicity and interrelation of infrastructure systems and percolation theory is the most common approach to quantify the robustness of such networks. This survey systematically reviews literature published between 2010 and 2021, on applying percolation theory to assess the robustness of infrastructure systems modeled as multilayer networks. We discussed all network properties applied to build infrastructure models. Among all properties, interdependency strength and communities were the most common network property whilst very few studies considered realistic attributes of infrastructure systems such as directed links and feedback conditions. The review highlights that the properties produced approximately similar model outcomes, in terms of detecting improvement or deterioration in the robustness of multilayer infrastructure networks, with few exceptions. Most of the studies focused on highly simplified synthetic models rather than models built by real datasets. Thus, this review suggests analyzing multiple properties in a single model to assess whether they boost or weaken the impact of each other. In addition, the effect size of different properties on the robustness of infrastructure systems should be quantified. It can support the design and planning of robust infrastructure systems by arranging and prioritizing the most effective properties.

## I. INTRODUCTION

Near zero disruption in the operation of infrastructure systems is vital for the delivery and integrity of the essential goods and services [1]. Any interruption in the performance of critical infrastructure systems, such as water distribution, electric power, natural gas, communication, and transportation systems can cause economic loss and affect societal wellbeing [2]. As robustness is referred to as the ability of a system to withstand disturbances [3], robust infrastructure systems are essential for continued societal function and averted economic loss. During recent years, infrastructure systems have evolved into complex systems with increasing interdependency [4], [5] mostly as the cyber-physical systems where communication function of telecommunication systems integrated into cyber interdependence of other infrastructure sectors like the power grid, water supply, and transportation systems [6]–[8]. On the other hand, the other types of interdependency including functional, geographical, and logical [5], [9] among infrastructure sectors exhibits complex relationships since the performance of a sector can be dependent on the proper functionality of other sectors. Thereby, a failure in one sector may cause damages to other interdependent sectors, leading to cascading failures spreading back-and-forth through the sectors and resulting in catastrophic consequences [10]. In this regard,

The associate editor coordinating the review of this manuscript and approving it for publication was Fung Po Tso.

the network-based models have been among the most popular approaches to explore the behavior of these complex systems under disturbances [11]. In this regard, multilayer networks have been developed which constitute layers of the infrastructure systems with nodes paired by links through different layers showing interdependency and interconnectivity [12].

The robustness in networks is defined as the ability to sustain a sequential nodes/links removal, as the simulation of arising and spreading disturbances in real-world systems [13]. During a disturbance, nodes/links are removed from a networked system due to five major reasons including imposing initial failures directly, failures of their couplings, becoming isolated, being part of the smaller emerging sub-networks that remained after stopping the cascade process, or locating within a certain distance from the other removed nodes/links [13]–[15]. The initial failures can be imposed by different types of attacks from random, to localized, and targeted (complementary information is provided in section III.N.1). The giant connected component is the largest sub-network of connected nodes/links remained after the cascade of failures through the whole network [15]. In this regard, the robustness of the networked system is measured based on the size of the giant connected component remained after the disturbance. The higher the number of remained nodes belonging to the giant connected component, the network is more robust [14].

These definitions are close to the theoretical approach of the percolation theory to study the behavior of networks when some nodes and links are designated as removed [16], [17]. It can be the reason why the percolation theory is the most popular in studying and analyzing the robustness of networked systems [18]. The percolation theory is to study the behavior of networks in terms of the critical percolation threshold and the size of the giant component, when some nodes and links are assigned as removed [16], [17]. In percolation theory, the removal fraction of nodes/links is tuned increasingly from zero to a certain critical threshold in which the state of the network shifts from connected to disconnected according to the decreasing size of the giant connected components [19]. Below the critical threshold, there is no giant connected component, whereas above the critical threshold a giant connected component exists [14].

However, many studies related to percolation theory have been focused on single-layer networks while real-world systems, including infrastructure systems, are composed of multilayer networks [20]. Applying percolation theory for analyzing multilayer networks was initiated by Buldyrev *et al.* [14], [16], [20], [21]. They utilized the percolation approach to study the robustness in a 2-layer network by defining two types of links; (i) connectivity links that connect nodes within layers and (ii) dependency links that connect nodes between layers. While they fixed the number of dependency links in their model, they varied the degree of nodes based on the number of connectivity links and found that the robustness decreases by having broader degree distribution in layers. Since then, the application of percolation theory for studying and analyzing the robustness of multilayer networks has been expanded, especially for evaluating the effects of different network properties [20], [22].

Parshani *et al.* [23] defined a "coupling strength measure" between two layers of a multilayer network based on the number of dependency links. They indicated that as the number of dependency links increases, the coupling strength increases, and ultimately, the vulnerability of multilayer networks increases. Cellai *et al.* [24] showed that links overlap among layers can improve the robustness of multilayer networks where two nodes are connected within all layers. In comparison to Shao *et al.* [25] that developed a theoretical framework by randomly interconnected nodes of two layers, some other researchers proposed frameworks with correlated interconnection [26]–[28]. It is concluded that nodes that are interconnected in a 2-layer network according to their degree correlation can improve the robustness of multilayer networks. Regarding that multilayer networks can be considered as n-interdependent layers, Gao *et al.* [29]– [30] developed a general framework to study the percolation of the networks of networks (NoN) with different structures of layers including random regular, tree-like, star-like, and loop-like. In contrast to other studies defining random removal of nodes to simulate random damages in real systems, [31], [32] were among the first researchers that studied the percolation of multilayer networks under targeted removal of nodes [20], [21]. These are some initial studies that were followed by more extensive research to study the effect of different network properties on the robustness of multilayer networks.

In spite of different studies that focused on capturing the effect of different network properties on the robustness of multilayer networks, the question is which of these properties have been used to represent different characteristics of infrastructure systems. This study provides a systematic review to gather all network properties used to build multilayer network models of infrastructure systems and discuss their impact on the robustness of these systems. We also categorize different mechanisms of nodes and links removal defined to simulate real-world failure propagations. The scope of this study is narrowed to the application of percolation theory. According to the findings, the effective network properties make approximately similar impacts which should be considered in designing and planning the future infrastructure systems and improving the structure of existing systems. Selected papers merely focused on assessing the effect of different network properties on the robustness of infrastructure while there is a lack of comparing the effect of different nodes/links removal mechanisms on the robustness.

The rest of the article is organized as follows. Section 2 explains the methodology used for the collection and selection of papers. Section 3 represents the contents of the selected papers by categorizing network properties. Section 4 contains extensive analysis and comparison between network properties and their effects. Section 5 presents our conclusion and thoughts about future works.

## II. METHODOLOGY

This study is a systematic review using Kitchenham's protocol [33] as a guideline. In the following sections, we explain how we planned and conducted the review to select papers and extract data.

### A. PLANNING THE REVIEW

This review paper aims to identify and describe how network properties, used to characterize infrastructure systems, influence the robustness and vulnerability of such systems. The focus is on answering the following questions: (i) what are the network properties used to model infrastructure systems as multilayer networks, specifically in percolation-based studies? (ii) how these defined properties affect the robustness of the created models? (iii) do these properties lead to distinct outcomes under different percolation mechanisms? (iv) which percolation mechanism has a more destructive effect? Since the focus is on network properties that are applicable for different types of infrastructure systems, we excluded papers containing models which were specialized and dedicated to a specific type of infrastructure.

Four databases including Scopus, Web of Science, Google Scholars, and Cochran were searched to find other similar review papers based on the applied inclusion/exclusion criteria of this study. Only five review articles turned up [16], [20]–[22], [34], yet none of them was a systematic review. In more detail, they did not contain the paper selection criteria and data processing. On the other hand, they considered other real-world systems such as biological, metabolic, and social systems in addition to the infrastructure systems. Therefore, this study is the first systematic review paper that gathered all network properties applied to build multilayer network models that represent infrastructure systems.

### B. CONDUCTING THE REVIEW

The intersection of three research areas defined the search query for this study; (i) multilayer networks, (ii) infrastructure systems, and (iii) percolation theory. The intersection was identified by linking three search strings listed in Table 1 with "AND". Adapted from Boccaletti *et al.* [35], in this study, a multilayer network is defined as "a pair M = (G, C) where G represents different layers of M, made by directed or undirected, weighted or unweighted graphs, and C is the set of interconnections between nodes belonging to different layers". It is a general framework for multilayer networks which can include other types of multilayer networks like interdependent networks, multiplex networks, networks of networks, and so on. Since the literature on multilayer networks is messy [36] and defined network structures in different studies are not consistent, providing definitions for different types of multilayer networks is out of the scope of this study. This study only focuses on network properties used to model infrastructure systems. Keywords in the search string related to multilayer networks were identified from [36] which were connected with "OR" to include all appropriate alternatives. To make it more inclusive, we also added similar word combinations made by the system instead of the network. Two single keywords used as the second and third search strings were sufficient and efficient to show the most relevant papers in the related areas.

We searched relevant peer-reviewed papers published after 2010 in Scopus and Web of Science. To find peer-reviewed papers, the document type and the source type were chosen as "Article" and "Journal", respectively. To collect the most relevant papers, the search query was limited to titles, abstracts, and keywords of articles. The language was limited to English. Table 2 specifies all applied inclusion and exclusion criteria.

**TABLE 1.** Search query.

| | | |
|---|---|---|
| Search query | Search string 1 | "complex system*" OR "inter$dependent system*" OR "system of systems" OR "inter$linked network*" OR "inter$related network*" OR "inter$active network*" OR "inter$connected network*" OR "hierarchical network*" OR "inter$dependent network*" OR "complex network*" OR "multi$layer network*" OR "multiple network*" OR "multiplex network*" OR "Multi$variate network*" OR "Multi$network*" OR "Multi$relational network*" OR "Multi$dimensional network*" OR "Multi$slice network*" OR "Multi$type network*" OR "network of networks" OR "coupled network" |
| | | AND |
| | Search string 2 | Infrastructur* |
| | | AND |
| | Search string 3 | Percolat* |

**TABLE 2.** Inclusion and exclusion criteria.

| Inclusion criteria |
|---|
| a) The paper is written in English |
| b) The paper is a primary study |
| c) The paper is peer-reviewed article published in a journal |
| d) The paper is published after 2010 |
| e) The paper is found on Scopus |
| f) The paper contains the search query in its title, abstract, and keywords |
| g) The paper applied percolation theory to study the robustness in infrastructure systems. |
| h) The paper studies interdependencies in a multilayer network. |
| i) The paper studies the interdependencies in a single-layer network by defining communities of nodes. |
| Exclusion criteria |
| a) The paper is not available online |
| b) The paper does not define its own network model |
| c) The paper is about single-layer networks |
| d) The paper contains models which are specialized and dedicated to a specific type of infrastructure. |

**TABLE 3.** Detailed information about extracted data in terms of the properties of networked models and their categories.

| Property | Synthetic/reality-based networks | | Network type | | | | | |
|---|---|---|---|---|---|---|---|---|
| Categories | Synthetic | Real | Erdös-Rényi (ER) | Scale-free (SF) | Lattice | Random regular (RR) | Small world (SW) | Real |
| Num. | 29 | 2 | 25 | 13 | 5 | 3 | 1 | 2 |
| Ref. | [24,39-41,44,47-55,59-61,64-66,73,76,79-82,90,101-103] | [37-38] | [24,40-41,44,47-49,51-55,59-60,64-66,73,76,81-82,90,101-103] | [39-41,49,51-55,59,64,90,102-103] | [39,50,61,80,102] | [59,79,102] | [41] | [37-38] |
| Property | Form of interdependence | | | | Attack type | | | |
| Categories | Full | Partial | Correlated | Tolerated | Random | Targeted | Localized | Probabilistic |
| Num. | 18 | 14 | 5 | 3 | 19 | 10 | 4 | 1 |
| Ref. | [24,37,39-41,47,49-52,59,61,73,76,79,82,90,103] | [38,44,48,53-55,60,64-66,80,81,101-102] | [39,50,49,51,79] | [53-55] | [24,37,39,47,48,50,52-55,59,66,76,79,80,82,90,102-103] | [38,40,41,49,51,60,64,65,73,81,101] | [37,39,44,51] | [37] |
| Property | Feedback conditions | | Spatiality | | Temporality | | Overlap | |
| Categories | Without | With | Without | With | Without | With | Without | With |
| Num. | 30 | 1 | 25 | 6 | 24 | 7 | 28 | 3 |
| Ref. | [24,38-41,44,53,55,59-66,73,76,79-82,90,101-103] | [37] | [24,40,41,44,47-49,51-55,59,60,64-66,73,76,79,81,82,90,101-103] | [37-39,50,61,80] | [24,37,39-41,44,47-48,51-55,59-61,73,76,79-82,90,101-103] | [38,49,50,53,64-66] | [37-41,44,47-49,51-55,59-61,64-66,73,76,79-81,90,101-103] | [24,50,82] |
| Property | Multilayer structure | | | | | | Directed links | |
| Categories | 2-layer | Tree-like | Star-like | RR | Loop-like | Single layer | Without | With |
| Num. | 21 | 9 | 3 | 2 | 1 | 2 | 30 | 1 |
| Ref. | [24,37,39,41-44,46-48,50,53,54,61,65,66,79-81,90,101-103] | [40,48,51,55,60,64,73,76,82] | [44,47,64] | [40,44] | [60] | [52,59] | [24,37-41,44,46-55,59,61-66,73,76,79-82,90,101-103] | [59] |
| Property | Topological/functional/ dynamical properties | | | Communities | | | Same nodes in layers | |
| Categories | Topological | Functional | Dynamical | Dependency gro. | Module | Cluster | Without | With |
| Num. | 27 | 2 | 2 | 4 | 2 | 2 | 20 | 11 |
| Ref. | [24,37,39-41,44,46-48,50-55,59-65,73,76,79-82,90,101-103] | [38,49] | [53,66] | [49,51,52,59] | [40,60] | [38,44] | [37,40,44,48,49,52,59,60,64-66,79-81,90,101-103] | [24,41,47,50,53-55,73,76,82] |
| Property | Percolation models | | Start point of failure | | Removal mechanisms | | Failure propagation through layers | |
| Categories | Site | Bond | In one of the layers | In all layers | Classic | Non-classic | Within and between layers | Within layers first |
| Num. | 28 | 3 | 25 | 6 | 27 | 4 | 28 | 3 |
| Ref. | [24,38-41,44,48-55,59-66,73,76,79-82,90,101-103] | [37,47,52] | [24,37,38,40,41,47-55,60,65,66,76,79-82,90,101-103] | [39,44,59,61,64,73] | [24,37-40,44,47,49-55,59,60,64-66,73,79-82,90,101-103] | [41,48,52,76] | [24,39-41,44,46-55,59-66,73,76,79-82,90,101-103] | [37-38,49] |

Once the initial collection of papers was done and duplications were removed, all collected papers were read in two phases to ensure the final group of selected papers meets the defined criteria; (i) title and abstract, (ii) the full article. As specified in Table 2, single-layer networks or single infrastructure systems were excluded. In addition, papers that only reviewed the results of other papers and did not build their own multilayer models were excluded.

In the next stage, in-depth reading and analyzing was performed to extract and classify data obtained from the selected papers. Data was classified into three main groups; 1- data related to different types of removal mechanisms and spreading the failure, 2- data related to different network properties defined for multilayer models, and 3- obtained results. More details about the extracted data and the groups they belong can be found in Table 3.

## III. RESULTS

The literature search identified 46 publications. Application of inclusion and exclusion criteria (Table 2) identified 31 papers as relevant to the scope of this review. The 15 excluded papers consisted of 8 papers that built models for single-layer networks, 3 papers which mostly contained general discussions without building their own models, and 4 papers modelled specific type of infrastructure (2 on the telecommunication system, 1 on the energy system, and 1 on the transportation system). Among 31 selected papers, there were two papers that built their models considering a single-layer network. However, their single-layer models contained dependency groups to evaluate interdependencies and interconnectivities. Thus, they were included in the relevant studies. According to the results, more than 60% of selected papers were published after 2018 which shows the growing trend of such studies. The detailed information compiled from included studies is presented in Table 3.

### A. SYNTHETIC/REALITY-BASED MODELS
Two studies used real datasets to build network models in which nodes were distributed in layers based on their real spatial locations. Degree, load, and capacity of nodes as well as direction, type, and the number of links were extracted from real datasets [37]- [38]. The rest of the papers applied synthetic networks to mimic different aspects of real-world systems as discussed in the following sections.

### B. NETWORK TYPE
Synthetic networks including Erdös-Rényi network (ER), scale-free network (SF), random regular network (RR), square lattice network (SL), and small world network (SW) were applied in the papers to define the configuration of nodes and links within layers. In [39], SF and SL were combined so that all nodes of two SF networks were randomly located in a 2D square plane with a defined side length.

In almost all but two of the synthetic network models, all layers were made by the same type of synthetic networks. But the two exceptions were [40] in which researchers built a hierarchal multilayer network of n layers where the top layer was ER and the lower layers were SF networks; and [41] in which researchers examined 2-layer networks composed of different combinations of ER-SF, ER-SW, and SF-SW networks. In [40], each layer represents a specific sector in infrastructure systems and each hierarchical level represents a geospatial scale as a state, cities, and neighborhoods from top to down. In [41], the multilayer model represent interdependency between different configurations of the communication network and the power grid.

By contrast to these synthetic models, [37], [38] built their models based on the real datasets. Thus, the location of nodes and arrangement of links in each layer were based on real-world cases of power/gas networks in [37], and road/sewer networks in [38]. All of these models can be grouped into five categories as illustrated in Figure 1; (i) models in which all layers are made by the same synthetic networks, (ii) models in which layers are formed based on different synthetic networks, (iii) models in which layers are built with different real datasets, (iv) models in which layers are composed of synthetic and real networks, (v) models in which all layers are made by the same real networks. The last two models were discussed in [17].

### C. NUMBER OF LAYERS
Among the included papers, 60% of the models were built as a 2-layer network. 35% of papers worked on multilayer networks of n layers. Two papers defined interdependencies and interconnectivities in single-layer networks by dividing nodes into different groups.

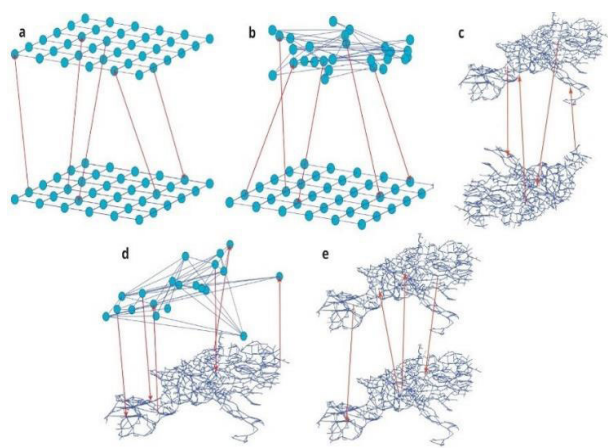

**FIGURE 1.** Five schematic diagrams to show multilayer network models in which; a) all layers are made by the same synthetic networks, b) layers are made by different synthetic networks, c) layers are made by different real networks, d) layers are made by synthetic and real networks, e) all layers are made by the same real networks (adapted from [33]).

### D. NoN-STRUCTURES
Multilayer models contained more than two layers were built according to the following Network of Networks structures (NoN) as illustrated in Figure 2; tree-like, random regular, star-like, and loop-like. Lattice and chain-like [17], [21] are also among NoN-structures which were not considered in the collected papers. These structures can represent how different cities are connected through their infrastructures [43]. They can also show the connection between different infrastructure sectors [44]. While Kivela et al. [36] declared the difference between interdependent networks and NoNs as links connecting two layers do not necessarily indicate dependency relations, most of the included studies in this review considered dependency relationships between layers in NoN-structures.

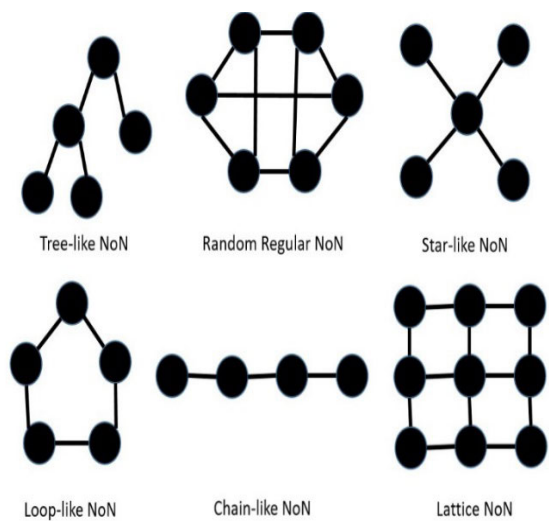

**FIGURE 2.** Schematic diagrams of different NoN-structures in which each node represents a network layer.

### E. FEEDBACK CONDITIONS

Except for one paper [37], all other papers considered no-feedback conditions in which network A depends on network B, then each node in A depends on one and only one node in B. As shown in Figure 3, in [37], feedback conditions were considered between schematic networks of road and sewer systems since there might be multiple sewer lines underneath one road link in which failure of any of them can lead to the road link closure [37]. However, the effect of the feedback conditions on the robustness of the multilayer network was not explained in this paper. Many real-world networks including infrastructure systems rarely appear in isolation, but always are coupled together, usually with feedback conditions [45]. [45], [46] assessed the effect of feedback conditions (also called feedback dependency links) concluding that they can make multilayer networks more vulnerable.

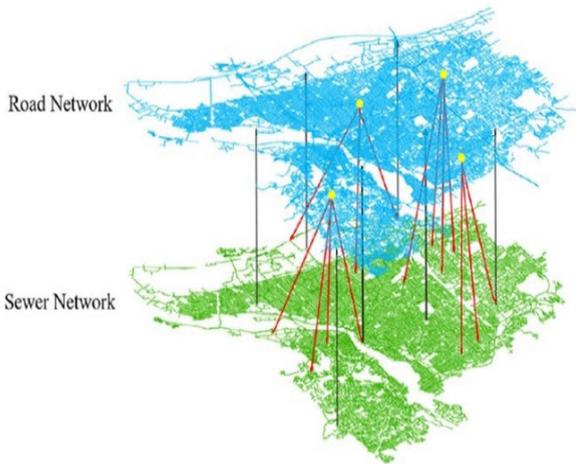

**FIGURE 3.** Schematic illustration of feedback conditions by red lines and no-feedback conditions by black lines between road and sewer networks [28].

### F. FORM OF INTERDEPENDENCE

We found four different forms of interdependency defined in the collected papers; full, partial, tolerated, and correlated (Figure 4). In full interdependence, every node in Layer A is connected to only one of the nodes in layer B. In this case, all layers have the same number of nodes like a multi-modal transportation linking multiple cities [47].

In partial interdependence, there are some nodes in layer A which are connected to some of the nodes in layer B. In this case, the higher the number of interconnected nodes, the interdependency is stronger. These two cases can be expanded to n-layer networks as well. As Gross *et al.* [48] explained partial interdependence is more realistic where additional resources are needed at a node to accommodate dependency links such as international airports equipped with longer runways for transoceanic flights and power stations equipped with additional infrastructures for transferring large loads to long distances.

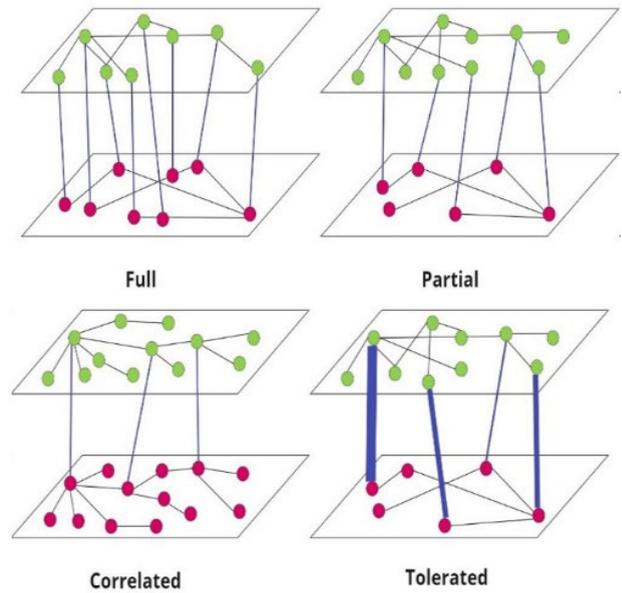

**FIGURE 4.** Different forms of interdependence patterns between layers. In full interdependence, there is a link between two nodes of different layers. In partial interdependence, only a fraction of random nodes is interconnected. In correlated interdependence, nodes are interconnected based on their degree. In tolerated interdependence, a different range of strengths is defined for dependency links.

The strength of interdependence can also be defined as the tolerance parameter which represents the tolerance of nodes in layer A to the failure of their interdependent nodes in other layers. In this case, some nodes with weaker interdependency are more resistant to damage of their interdependent neighbors while some nodes with stronger interdependency show low tolerance because they are vulnerable to the failure of their interdependent nodes. As Liu *et al.* [47] explained, in real-world systems, a failed node does not destroy all of its neighbors completely. For example, when an airport is shut down, it is not likely that all trips via cars, buses, trains, and ferries will be affected equally leading to removing all nodes and links. The failure in air travel can increase travel demand in each interdependent mode but they may have different capacities to support increased demand. In addition, the amount of added demand for each mode is not equal since passengers may choose one over the others. Thus, the failure of a node in one layer cannot necessarily cause the loss of all interconnected nodes and links in other layers.

Since real interdependent systems are usually not randomly connected, there is correlated interdependence among defined models. For example, worldwide ports tend to couple to international airports. Similarly, it is much more common that a central telecommunication center depends on a central power station [26]. There are two linking patterns in correlated interdependence; assortative and disassortative [39], [49].

In assortative patterns, nodes in different layers are interlinked by descending ranking of their degree or centrality.

In disassortative patterns, nodes with the higher degree or centrality in layer A tend to connect to nodes with the lower degree or centrality in other layers. The correlated interdependence is one of the two parameters which were introduced by Parshani et al. [26] to assess inter-similarity between layers in multilayer networks. The other one is the inter-clustering coefficient which evaluates how many of the neighbors of node i in layer A depend on neighbors of node j in layer B when i and j are interconnected. The effect of inter-similarity in terms of the inter-clustering coefficient was not studied in the selected papers of this paper.

Some studies only considered one of these interdependency forms in their models while some others studied how the combinations of these forms affect the robustness of multilayer networks. [39], [49]–[52] built their models with the combination of full and correlated interdependence. [53], [55] considered partially tolerated interdependence among layers. As Li et al. [16] mentioned, the combination of tolerated and correlated interdependency was studied in [56] which was not among the papers collected in this study. In this case, different tolerances were assigned to dependency links according to their degree correlation. So, nodes with higher degrees were more resistant to damage.

### G. TOPOLOGICAL/FUNCTIONAL/DYNAMICAL PROPERTIES

Studying the effect of interdependency in multilayer networks can be divided into three categories based on the nature of network properties applied in the models; topological, functional, and dynamical. More than 85% of the collected papers focused on the effect of topological properties such as number of layers, number of interconnected nodes, size of communities, and configuration of linking patterns. Functional properties in multilayer models of infrastructure systems were defined as the load and capacity of nodes and links used to produce and distribute a continuous flow of goods and services within and between the layers [57]. As Wang et al. [58] discussed the functionality of layers is dependent on the steady flow coming from other layers. In this case, each node has a certain initial load and capacity. When the network is damaged, the load of failed nodes will move to other nodes and push other nodes beyond their capacity limit and cause cascading failure [49].

As Danziger et al. [53] explained dynamical properties were defined as a node's local state (order or disorder) at a certain time which is dependent on the local state of its neighbors within and between layers. The overall state of nodes in a layer reflects the collective behavior of the layer. The varying state of nodes in a period of time can be studied under the influence of dynamic processes like epidemic spreading, synchronization, diffusion, and game theory.

### H. PRESENCE OF COMMUNITIES

In 8 studies, layers contained communities of nodes including dependency groups [49]–[51], [59], clusters [38], [44], and modules [40], [60]. Dependency groups were made by randomly dividing all nodes in a layer into the same size groups without overlapping. While [16] mentioned that some studies also worked on dependency groups with overlapping nodes, it was not the case among the selected papers in this study. According to the provided descriptions, if at least one of the nodes in a group fails, the whole group is removed from the network. However, there were some differences between the models containing dependency groups. Two studies evaluated dependency groups in a single layer model [52], [59] while the two others made 2-layer models [49], [51]. In 2-layer models, only one of the layers contained dependency groups and the other one was made by individual nodes. In [51], all nodes in a dependency group are treated as a super-node (Figure 5a). Two super-nodes in layer A are connected with a connectivity link when there is at least one link between their nodes. Each super-node in layer A is connected to an individual node in layer B with a dependency link. In [49], every node in the dependency groups can be connected with a node in the same layer or the second layer (Figure 5b). As an example, a substation in a power network distributes electricity needed in gas compressors, storage regulators, and control systems which are located in a specific area [38]. Another example is a Wi-Fi network that provides connections for devices within a determined radius [49].

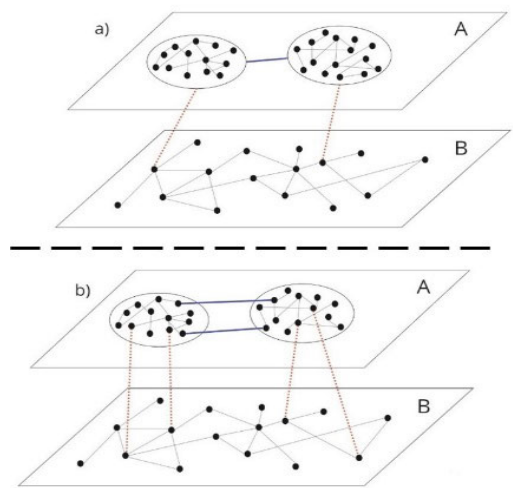

**FIGURE 5.** Different linking patterns among dependency groups where they perform as super-nodes in (a) and not super-nodes in (b).

Clusters were defined as densely connected communities of nodes in a layer with varying sizes that can have one or more links to other clusters within and between layers like where infrastructure components are distributed in close proximity [44]. Modules were groups of nodes in layers as each module has its specific feature (according to its specific function [60] or geographic location [40]). For example, in Figure 6, each layer contains three modules with three different colors where layers depict cities and each color represent specific types of infrastructure. In this case, dependency links between layers are restricted such that they only connect nodes of the same modules. Connectivity links within layers connect nodes in different modules. The major difference between modules and dependency groups is that modules can be interconnected between layers only when they belong to the same module while dependency groups do not have such a restriction.

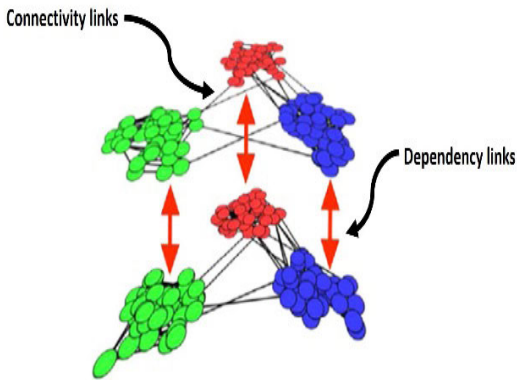

**FIGURE 6.** Dependency links are between the same modules (same color) in different layers and connectivity links are between different modules (different color) in the same layer [60].

### I. SPATIALITY
In multilayer networks, spatial-related data is an important factor influencing their robustness [39]. Vaknin *et al.* [61] declared that spatially embedded networks are more vulnerable than non-embedded networks. In spatially embedded networks, nodes and links are constrained in some inflexible geometry since they have the specific location with a certain distance from the rest of the network, specific number, size, capacity, and length which cannot be changed easily as it can be very costly and unfeasible [62]. In this way, spatiality was studied based on three aspects including; spatial location of nodes, length of links, and co-location of networks. Embedding nodes in specific locations is a fundamental quantity to characterize the structure of infrastructure systems and impose some spatial constraints for example because of geographical and topographical features [42]. On the other hand, different distributions of link length reflect different characteristics in infrastructure systems like highways, transmission lines, communication lines and pipelines [20], [39] where the vulnerability of the network depends on the critical length of links [63]. Co-location is another spatial-related feature implying that two networks have an effect on each other due to locating at a close distance. For example, water pipe leaks can damage road foundations since water and wastewater infrastructure are often buried underneath roads [37].

### J. TEMPORALITY
Among 31 selected papers, only 7 of them considered temporality in their multilayer models. The temporal properties defined in the papers can be grouped into two categories including; the number of time steps in which failures propagate across the network and the duration of time spans in which the network spends in each percolation phase. The first group is related to the changing size of the removed and remained parts or the failed and functional parts of the network during the time, from the initiation of the removal to the end of the cascading process. For example, [50], [64], [65] topologically assessed the fraction of removed nodes in different time steps. Similarly, [49], [66] evaluated the state of the failed and functional nodes at different time steps in terms of their capacity and load. In a different way, Wang *et al.* [38] estimated the size of the removed nodes under several targeted attacks during a period of time while others only defined one initial attack at the beginning of the removal process. The second group is about how long each percolation phase takes. Danziger *et al.* [53] displayed the duration that the network spends at each percolation phase; from the ordered phase (connected) to the transition phase and then the disordered (disconnected) phase. Section N in the following provides more information about different percolation phases.

### K. OVERLAP
Cellai *et al.* [24] in 2013 established the first study that considered percolation in multilayer networks with overlap [20]. As node i and j exist in some or all layers of the multilayer networks, the link (i,j) is overlapped if it connects node i and j within more than one layer. In a fully overlapped multilayer network, two nodes are connected within all layers like two cities as two nodes in the multilayer transportation network in which they are connected by the road network, railway network, and flight network.

### L. NODES
All studies consider the same number of nodes in all layers except for two real cases. While 18 studies considered different nodes in layers, 11 studies examined the existence of the same nodes in all layers; like electrical substations as nodes which are connected by underground and overhead lines to transmit electrical power as well as optical fibers to communicate and remotely control the substations. In this case, as at least one node has a counterpart in some other layers, it is called multiplex networks [67]. Some studies defined weights for nodes based on the specifications of their models. For example, degree and centrality of nodes, capacity and load of nodes, and tolerance parameter were defined to evaluate the mechanism of targeted attacks, the effect of functional properties, and strength of interdependency, respectively.

### M. LINKS
Links can be categorized if they are internal or external, unidirectional or bidirectional, weighted or not weighted. As mentioned before, there are two types of links in multilayer networks [14]; connectivity links that connect nodes within layers (internally) and dependency links that connect nodes between layers (externally).

Among all collected papers, only [60] built unidirectional dependency links between layers in loop-like NoN. All other papers considered bidirectional links within and between layers since they assumed that there is a two-way interdependence between networks [38]. For example, a power network produces electricity required for pumps, lift stations, and control systems in water network while water network provides water for cooling in power network [68]. In some studies [69], [70], bidirectional links are interpreted as interdependencies while unidirectional links are defined as dependencies. However, it is not the case in all studies; for example [14] called bidirectional connections as dependency links. [50], [61] evaluated the effect of link length as the link's weight for both connectivity and dependency links.

In [50], [61], weighted links were defined as the spatial length of connectivity and dependency links while [53]–[55] defined tolerate parameter as the weight of dependency links between layers. More information regarding the tolerance parameter can be found in Section F.

### N. PERCOLATION MODELS

Percolation theory is the simplest model displaying a phase transition [71] where the network undergoes transitions from the phase of a large connected component (functional network) to the phase of multiple dis-connected components (non-functional network) [24]. Percolation theory is applied to study how different networked systems are more robust or vulnerable, in terms of the critical percolation threshold and the size of the giant connected component. The giant connected component is the largest subnetwork of connected nodes/links that remained from the multilayer network after stopping the removal process. This part is considered as a "still functional part" of the initial network [17]. The higher the number of remained nodes belonging to the giant connected component, the network is more robust [14].

The critical percolation threshold is the fraction of removed nodes or links leading to the complete collapse of the network [22]. Below the critical threshold, there is no giant connected component, whereas above the critical threshold a giant connected component exists [14]. The robustness and vulnerability of multilayer networks can also be measured according to observing one of the two types of percolation phase transitions [72]. As illustrated in Figure 7, this transition in the size of the giant connected component can be abrupt/discontinuous at the critical threshold implying the vulnerability of the network since the first-order phase transition occurs. In contrast, the system is robust if the transition is continuous with a gentle slope. In this case, we have the second-order phase transition.

To model different disturbances in infrastructure systems, the removal process of nodes/links in the collected papers were defined based on different algorithms which can be explored by answering the following questions; a) how does it start?, b) which layer does it start from?, c) does it propagate through nodes or links?, d) how does it propagate within and between layers?, e) how does it form the giant connected

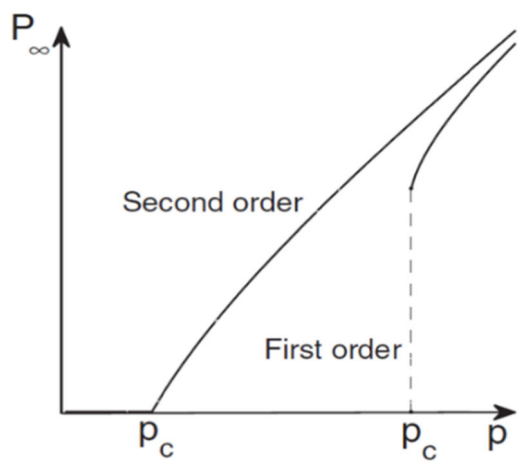

**FIGURE 7.** Schematic diagram to denote smooth transition in the second order percolation and abrupt transition in the first order percolation. $P_\infty$ is the fraction of nodes in the giant connected component, p is the fraction of remained nodes, and $p_c$ is the critical threshold [72].

component?. In the following section, we discuss more the given answers to these questions.

#### 1) ATTACK TYPES

To answer the first question, six types of attacks were identified in the papers as the simulation of real-world failures initiated at the beginning of the removal process; including random attacks, degree/centrality-based targeted attacks [38], [40], [49], [51], [64], [73], [101], dependency-based targeted attacks [40], [60], [65], oriented/focused localized attacks, and probabilistic attacks (Figure 8). In all types of attacks, each node/link is removed with probability 1-p. So, at the end of the removal process, p fraction of nodes remains. If remained nodes belong to the giant connected component, they can be considered as still functional part of the network. However, depending on the type of the attack, the first round of removal happens in different ways. In random attacks (also called random failures [74]), removing nodes is random to represent accidental and random damages (occur deliberately or not deliberately) in real-world systems (Figure 8a). Targeted attacks remove nodes based on their criticality in multilayer networks. Nodes with higher degree/centrality as well as nodes with dependency links play a more critical role such that they can be targets in malicious attacks. Oriented and focused localized attacks (Figure 8b and 8c) can represent damages caused by natural hazards like earthquakes and floods, respectively [74]. In this way, core/root nodes are picked randomly or trajectorially, depicting the center of the damage. Then, all nodes/links within the specified distance from the core are removed abruptly [61], or gradually (shell by shell) according to increasing distance [37], [39], [44]. Dong et al. [37] proposed a framework to model probabilistic attacks to analysis the system performance under post hazard scenarios. Probabilistic attacks represent the case in which some nodes/links have higher probability to be removed from the network. For example, roads close to water are more likely to liquefaction after the earthquake.

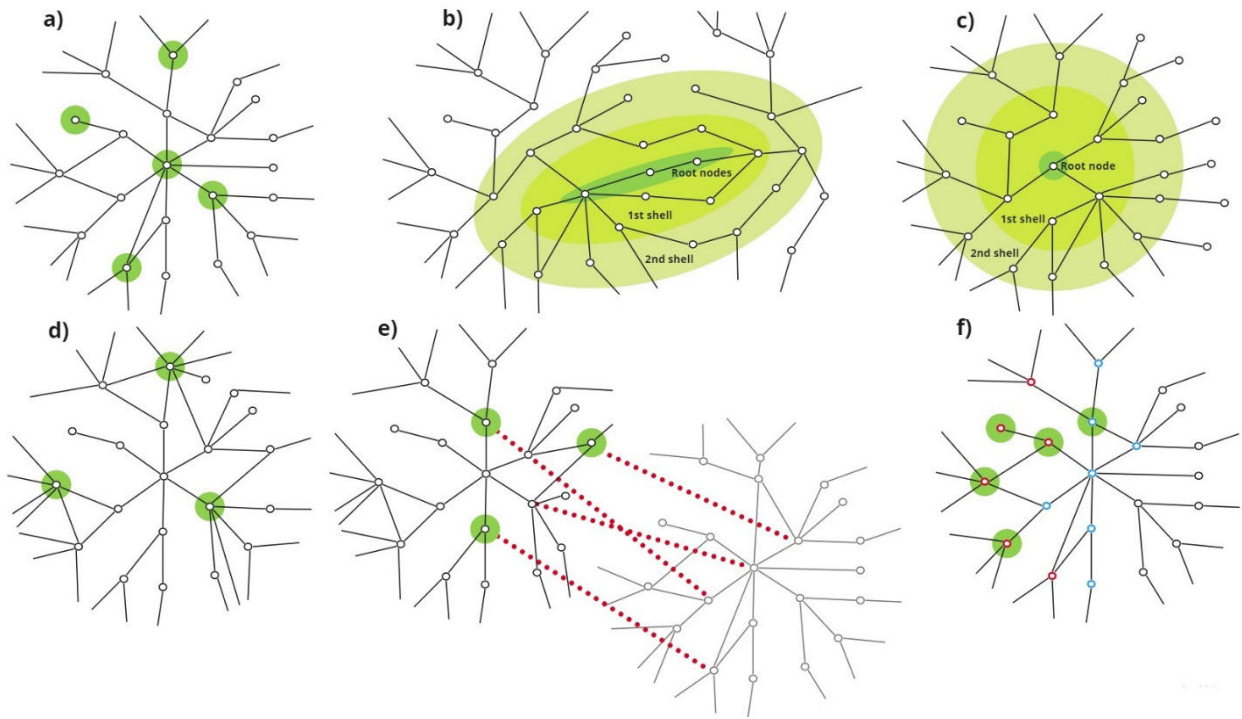

**FIGURE 8.** Schematic diagram to show the initiation of attacks in a layer; a) random attack in which some nodes are selected and removed randomly, b) oriented localized attack in which some nodes in the same trajectory are removed first and then the failures propagate sell by shell or abruptly, c) focused localized attack in which some individual nodes are removed randomly first and then the failures propagate sell by shell or abruptly, d) degree-based targeted attack in which nodes with a higher degree are removed first, e) dependency-based targeted attack in which nodes with dependency links are removed first (red dot-lines represent dependency links), f) probabilistic attack in which nodes with higher probability are removed first (from left to right, red nodes have a higher probability, blue nodes have moderate probability, and black nodes have low probability to be removed).

### 2) PERCOLATION MECHANISMS OF FAILURE

To answer Question b, different mechanisms related to how failures propagate within and between layers during the percolation process were identified. In 25 papers, it was indicated that the damage started from one of the layers while in six papers, the damage was triggered in all layers simultaneously [39], [44], [61]–[64], [73] where there are replicas of a node in other layers [61], [73] or in the case of localized and targeted attacks with a concurrent damage to different infrastructure sectors [44], [64]. To answer Question d, three articles [37], [38], [49] indicated that the number of removed nodes/links needs to reach the specific level within the layer to spread to the other layers (Figure 9b) such as a case where failure of a few low-load nodes in a gas network does not cause other nodes overload and failure [38]. In the other articles, it was specified that when a node with a dependency link fails, the failure can spread within and between layers at the same time (Figure 9a). Vaknin et al. [61] introduced the directed giant connected component for a fully interconnected multilayer network where a node is removed only if it is in contact with at least a removed node in upper layers (Figure 9c). So, the failure propagates from top to bottom layers like a case where a failure spreads from the asset layer to the component functionality layer in cyber-physical infrastructures [75].

To answer Question c about whether percolation propagates through nodes or links, two mechanisms exist in the papers; site percolation and bond percolation. Site percolation refers to the node-to-node failure propagation in which removal of a node leads to removal of the connected links and then removal of the coupling node. Bond percolation refers to the link-to-link failure propagation in which eliminating a link removes the two connected nodes. In site percolation, the giant connected component is identified as the largest group of connected nodes while in bond percolation, it is the largest group of connected links [71]. There are three papers [37], [47], [52] among the selected papers in this study that considered the bond percolation while the rest of the papers worked on the site percolation. Li et al. [16] declared that the bond percolation is considered less general than the site percolation due to the fact that the bond percolation can be reformulated as a site percolation. In general, the effects of node-to-link failure propagation on the robustness of a multilayer network have not been studied yet [47]. Figure 10 shows percolation on the square lattice, where different colors denote different subnetworks, forming during the removal process of nodes/links. The right-hand-side pictures contain giant connected components where the fraction of removed nodes/links is lower while the left-hand-side pictures denote the formation of infinite small subnetworks due to a larger fraction of removed nodes/links.

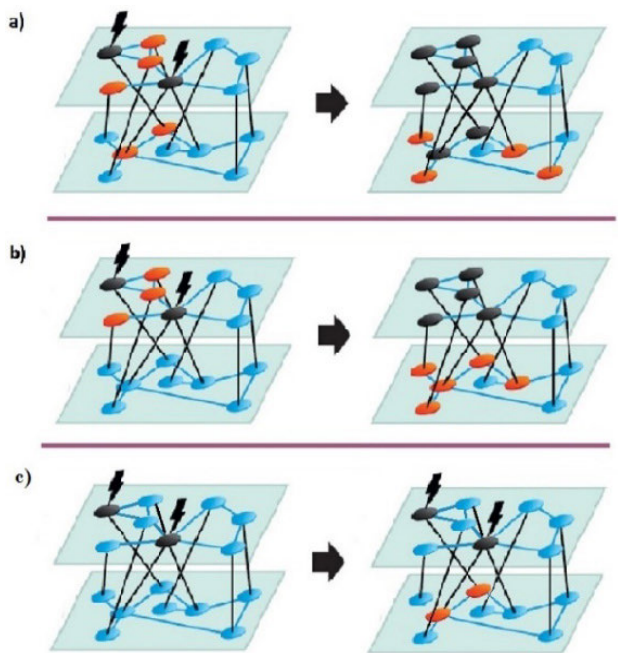

**FIGURE 9.** Different steps of failure propagation after removal of 2 nodes in a 2-layer network; a) failures propagate within and between the layers simultaneously, b) failures reach the specific level within the first layer and then spread to the other layers, c) failure spreads from the upper layers to lower layers. Black nodes represent removed nodes, red nodes represent affected nodes, and blue nodes are unaffected nodes (adopted from [22]).

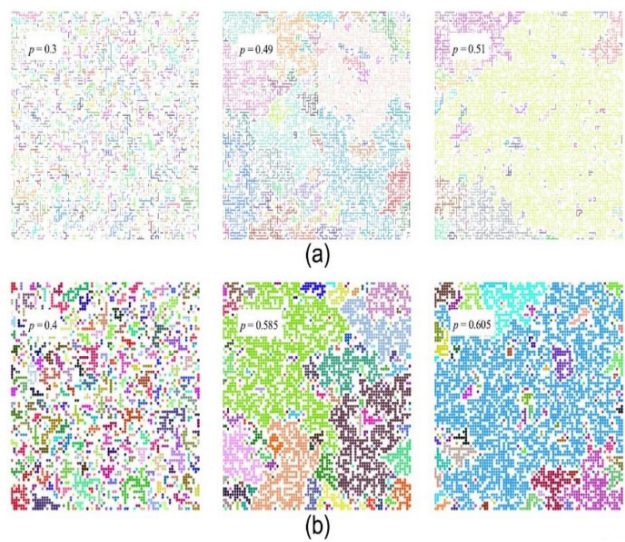

**FIGURE 10.** Different colors denote subnetworks in the bond percolation (a) and site percolation (b) on square lattice. P is the fraction of remained links/nodes [16].

Since different mechanisms of the node/link removal can results in a different giant connected component even with the same number of removals [16] it is important to identify them which can provide an answer for Question e. According to the classical percolation model, upon the initiation of the failure in a layer, it will propagate to other layers due to the interdependencies and interconnections in multilayer networks. This cascading failure proceeds back and forth between all layers until the multilayer network reaches a steady state when no further node/link removal occurs [39]. At this stage, we are left with the giant connected component [61] which is considered as the functional part of the remained network.

However, there are other recursive/iterative mechanisms defined for forming the giant connected component including [16]; k-core percolation, color-avoiding percolation, bootstrap percolation, greedy percolation, clique percolation, explosive percolation, etc. 27 of the selected papers considered classical percolation while the other 4 worked with other types of percolation models. Gross *et al.* [48] considered k-core percolation which is an iterative process of removing 1-p fraction of nodes, similar to the classical percolation. However, nodes with less than k neighbors are also considered as failed. Thus, nodes in the final giant component have at least k links to other remained nodes.

Bootstrap percolation which was considered in [76] is a recursive process in which every removed node can be recovered if it has at least k neighbors. Bootstrap percolation occurs when a subset of nodes with the specified size (called seeds) remains at the end of the percolation process [77]. According to the greedy framework which was applied by [41], removing articulation nodes causes the most damage to the network and reduces the size of the giant component. In more detail, removing articulation nodes maximizes the number of failed nodes and minimizes the connectivity of nodes across the network [78].

In color-avoiding percolation, used in [52], nodes/links are classified into groups based on their common vulnerability like buses that can be delayed during rush hours while undergrounds can be affected by technical problems. Each group gets its own color. In this regard, the giant component remaining after the percolation process is the intersection of nodes/links that always remain connected when each group color is removed iteratively.

## IV. DISCUSSION

While 29 collected papers defined different structures for their models, they produced approximately similar outcomes in terms of the effect of network properties on the robustness of infrastructure systems as multilayer networks. There are some exceptions mentioned in Table 4 which are discussed more in the following sections.

### A. THE EFFECT OF TOPOLOGICAL/FUNCTIONAL/DYNAMICAL PROPERTIES

According to the obtained results, stronger interdependence, in terms of either a greater number of dependency links (in full/partial interdependence) or lower tolerance (in tolerated interdependence) decreases the robustness of the multilayer network [44], [48], [54], [55], [64]–[66], [79]. In this

**TABLE 4.** Comparing the effect of different network properties on the robustness of infrastructure models.

| Network part | Network property | Positive/negative relationship with robustness* | Exceptions |
|---|---|---|---|
| Links | Average degree of nodes within layers | 🟢 | - |
| | Correlation of dependency links | 🟢 | For SF under degree-based targeted attacks |
| | Length of links | 🔴 | - |
| | Number of dependency links | 🔴 | - |
| | Number of fully overlapped links | 🟢 | - |
| | Number of links between modules within layers | 🟢 | - |
| | Number of links between dependency groups within layers | 🟠 | - |
| | Number of links between clusters within layers | ⚪ | - |
| | Presence of directed links | 🟠 | For SF networks |
| | Presence of feedback conditions | 🟠 | - |
| Nodes | Capacity of nodes | 🟢 | - |
| | Embedding nodes in space | 🔴 | - |
| | Inter-similar clustering coefficient | 🔵 | - |
| | Number of nodes in layers | ⚪ | - |
| | Number of overlap nodes between dependency groups | 🟠 | - |
| | Number of reinforced nodes | 🟢 | - |
| | Number of high load nodes in modules | ⚪ | - |
| | Number of high load nodes in dependency groups | ⚪ | - |
| | Number of high load nodes in clusters | 🔴 | - |
| | Strength of interdependency between nodes | 🔴 | - |
| Communities | Inter-similarities between interdependent communities | 🔵 | - |
| | Number of modules within layers | 🔴 | - |
| | Number of dependency groups within layers | 🔴 | - |
| | Number of clusters within layers | 🔴 | - |
| | Size of modules within layers | 🔴 | - |
| | Size of dependency groups within layers | 🔴 | For super-nodes |
| | Size of clusters within layers | 🔴 | - |
| Layers | Colocation of networks | 🔴 | - |
| | Number of layers | 🔴 | For large number of layers |

* The red box shows the negative relationship between the network property and robustness, included by the selected papers. The green box shows the positive relationship between the network property and robustness, included by the selected papers. The orange box shows the negative relationship between the network property and robustness, included by the review papers. The blue box shows the positive relationship between the network property and robustness, included by the review papers. The grey box shows the property which was not included in both groups of papers.

regard, Fan et al. [80] found the optimal fraction of interdependent nodes as r = 0.1. It suggests that if 10% of cities have interconnected flights, the network is the most robust to random attacks. The assortative correlation between interdependent layers can improve the robustness of multilayer networks in contrast to the disassortative correlation [39], [49], [51]. Nonetheless, [49], [51] showed that in the case of degree-based targeted attacks, correlated interdependence does not have an effect on the robustness. In general, the inter-similarity between networks regarding both the inter degree-degree correlation and the inter-clustering coefficient boosts the robustness of multilayer networks [26].

Studying dynamical properties showed that dynamic processes (like synchronization, epidemic spreading, diffusion, etc.) undergo discontinuity in partial interdependence while they go continuous in full interdependence [53]. For example, nodes synchronize abruptly in full interdependence and smoothly in partial interdependence.

Improving the robustness of multilayer networks in terms of the functional properties can be evaluated by enhancing nodes. In this case, increasing the capacity of nodes can increase the robustness of multilayer networks [38], [49]. In addition, areas with many high load nodes are more vulnerable so they should be given prioritized protection. [64], [81] found that there is a minimum density of randomly reinforced nodes among all nodes helping with eradicating catastrophic collapse. The universal density which they found is 0.1756 proved to be true for ER networks with any average degree and SF and RR networks with a large average degree. It was examined in 2-layer models and single-layer models with dependency groups. As an example, critical buildings in the power transmission grid, facing a sudden power outage, can employ backup facilities such as distributed generators. Reinforced internet ports, after cutting off their fiber links, can also use satellites or high-altitude platforms to exchange vital information [59].

### B. THE EFFECT OF OVERLAP

While full overlap increases the robustness of a multilayer network, partial overlap can make the network more vulnerable [24], [50], [59], [76]. Cellai et al. [82] demonstrated that for the same structure of multilayer network, the case in which all layers completely overlap is more robust. In [24], it was explained that the effect of partial overlap in multilayer networks with more than two layers is less predictable and it mostly depends on the structure of overlaps. For example, in a three-layer network the ratio of single overlaps (c1), double overlaps (c2), and triple overlaps (c3) play a critical role. When there are higher proportions of c2 and c3 than c1 then the network is more robust while with higher proportions of c3 and c1 than c2 the network is more vulnerable. Figure 11 illustrates the concept of c1, c2, and c3.

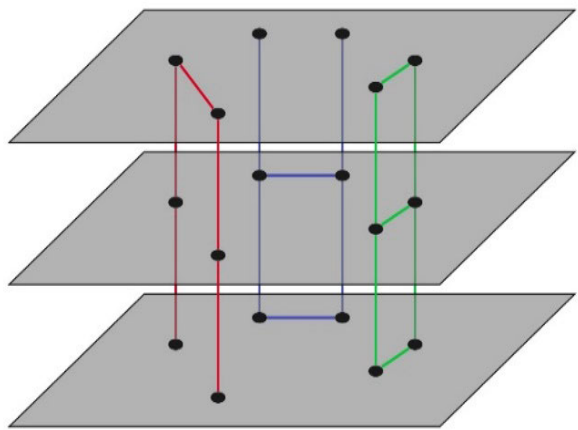

**FIGURE 11.** The red, blue, and green links represent single overlaps (c1), double overlaps (c2), and triple overlaps (c3) respectively in a three-layer network.

### C. THE EFFECT OF NODE DEGREE

Higher average degree of nodes within the layers enhances the robustness [38], [40], [47], [49], [55], [61]–[66]. In other words, more connections among nodes within layers result in a more robust multilayer network. If enough connections are available within a layer, in the case of losing interdependent links, the layer can continue to work [65]. In multilayer networks made by SF networks, the increase of average degree has a larger impact on the robustness in comparison to ER networks [49], [55].

### D. THE EFFECT OF COMMUNITIES

In general, multilayer networks with communities of nodes including dependency groups, clusters, or modules are more vulnerable than networks without communities [47], [52], [60]. Papers studying the effect of communities also reported that in the multilayer networks, more numbers and bigger sizes of communities can increase the vulnerability. [38], [44] indicated that denser clusters of nodes and links make the multilayer networks more vulnerable. In more detail, areas with the small number of nodes and links collapse after the failure of a few nodes but it does not cause a huge loss in the whole system. However, failures in areas with more nodes and links can cause significant performance drop in the whole system. While [40] showed that more and bigger modules can worsen the robustness of multilayer networks, Shekhtman et al. [60] reported that the size of the modules did not make a significant difference in their case. [44], [52], [59] declared that more and bigger dependency groups make multilayer models more vulnerable. However, there is one exception. Dong et al. [39] showed that if we treat all nodes of a dependency group as a super-node then bigger dependency groups result in more robust multilayer networks (refer to Figure 5).

Shekhtman et al [60] indicated that more connectivity links between different modules can make the multilayer networks more robust. However, such a study did not exist for dependency groups and clusters among included papers in this study. In this regard, [16] mentioned a study that worked on the effects of connectivity links exciting between dependency groups within the layer [83]. The study found that connectivity links between dependency groups have the same effect as dependency links between layers as they both make multilayer networks more vulnerable. They also discussed when these two types of links have nearly the same number, multilayer networks show more robustness.

Li et al. [21] indicated studies working on the effect of inter-similarity and overlapping between dependency groups which were not among the collected papers of this study. However, they can be the case in infrastructure systems as well, like different groups of electronic devices with a shared internet connection [28]. For example, [28], [84] showed that similarity between dependency groups in terms of their degree correlation, link overlapping, and the number of nodes can result in more robust networks. Similarly, [85] worked on dependency groups with shared nodes and proved that more than one shared node between two dependency groups makes the network more vulnerable.

In multilayer networks with modules under dependency-based targeted attacks, we should expect a phenomenon called multiple percolation transition where there is more than one jump in the percolation transition. As shown in Figure 12, the multiple transitions depict how failure propagates through the network and separates layers first and then modules. Shekhtman et al. [40] worked on modular multilayer network with hierarchical structure and demonstrated that the maximum number of jumps is the number of layers minus 1. There can be fewer jumps depending on other parameters like the average degree of nodes. In this regard, the lower the number of jumps, the multilayer network is more vulnerable.

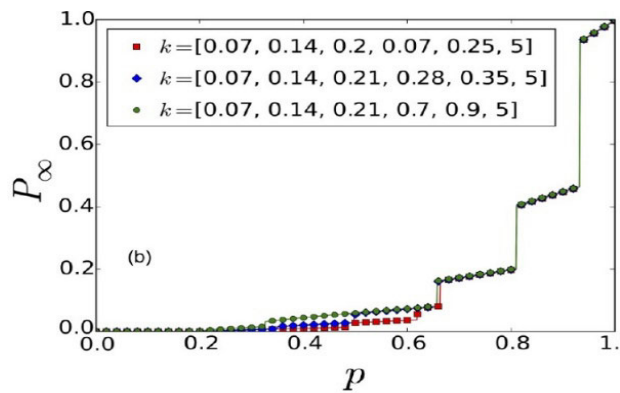

**FIGURE 12.** The multiple percolation transitions of cascading failure in a 6-layer model with the hierarchical structure of modules. $P_\infty$ is the fraction of nodes in the giant connected component, p is the fraction of remained nodes, and each value for k represents the average degree of nodes in each layer [40].

### E. THE EFFECT OF NUMBER OF LAYERS
According to the collected papers in this study, more layers have a devastating effect on the robustness of the multilayer network [44], [54], [59], [61], [75]. However, Radicchi et al. [86] showed that in a network with a large number of layers adding a new layer can boost the robustness since it can create redundant interdependencies among layers. For example, adding another flight operator between two large cities can increase the size of the system and provide a backup option.

### F. THE EFFECT OF SPATIALITY
The effect of spatiality can be assessed from three aspects; embedded nodes in space, length of links, and co-locations of networks. As [61] indicated, embedding nodes in Euclidean space makes multilayer networks more vulnerable than non-embedded networks. Three papers that examined the effect of different link lengths concluded that shorter link length (for both connectivity and dependency links) can improve the robustness of multilayer networks [39], [50], [61]. Co-locating of networks can also have destructive effects as a failure in one network results in a failure in the co-located network [37], [38].

### G. THE EFFECT OF NoN-STRUCTURES
As mentioned before, different NoN-structures were applied in the collected papers including tree-like, random regular, star-like, and loop-like. Based on the results, they do not have any effects on the robustness of multilayer networks. However, [47] introduced a new phenomenon in the star-like NoN called mixed percolation transitions, where the peripheral layers percolate first continuously and then the hub layer percolates discontinuously. In this case, we have a moderate value of the tolerance parameter. For the relatively large value of tolerance parameter, the multiple percolation transitions occur where peripheral and hub layer percolate continuously but in different orders. For the small value of the tolerance parameter, all layers percolate simultaneously and abruptly. Figure 13 illustrates different formats of percolation in a star-like NoN, including mixed percolation transitions.

### H. THE EFFECT OF LINKS DIRECTION
[87]–[89] indicated that directed networks are more vulnerable than undirected networks in the case of single-layer networks. Liu et al. [90] assessed the effect of directed links in a 2-layer model and demonstrated that in-degree and out-degree correlations between interdependent nodes increase the robustness of multilayer networks with SF networks but it has an opposite effect on multilayer networks with ER networks. However, the only paper that considered directed links in this study [60] did not discuss their effects on the robustness of multilayer networks.

### I. THE EFFECT OF PERCOLATION MODELS
#### 1) THE EFFECT OF ATTACK TYPES
Degree-based targeted attacks make multilayer networks with SF networks more vulnerable [51], [64]. [49], [51] showed that in the case of degree-based targeted attacks, correlated interdependence does not have an effect on the robustness. Only did one of the articles study SW networks and showed multilayer networks made with this network present the best results under degree-based targeted attacks among all other networks [41]. Dependency-based targeted attacks lead to a more stable network than random attacks [65]. In more detail, by removing even all of the interdependent nodes, each layer can still be functional if enough connectivity links are provided to keep the nodes connected [60], [65]. Random attacks increase the fragility of multilayer networks with ER networks [41], [51], [54], [64]. Correlated interdependence has better performance under random attacks while shorter links work better under localized attacks [39]. As [44], [61] discussed localized attacks have more destructive effects, particularly in multilayer networks with communities. Vaknin et al. [61] defined a minimum radius of damage in localized attacks which is needed to push the network to the complete collapse. They also introduced an approximated size for multilayer networks in which there is no correlation between attack size and network size. In other words, when the network is large enough, the collapse happens once the damage starts, regardless of the size of the localized attack.

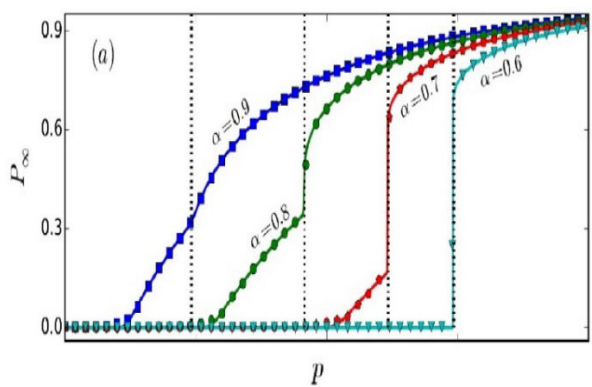

**FIGURE 13.** Displaying multiple percolation transitions (dark blue line showing two continuous transitions), mixed percolation transitions (green and red lines showing continuous and abrupt transitions together), and normal percolation transitions (light blue line showing an abrupt transition) in a star-like NoN of four layers. $\alpha$ is the tolerance parameter, $P_\infty$ is the fraction of nodes in the giant connected component and p is the fraction of remained [47].

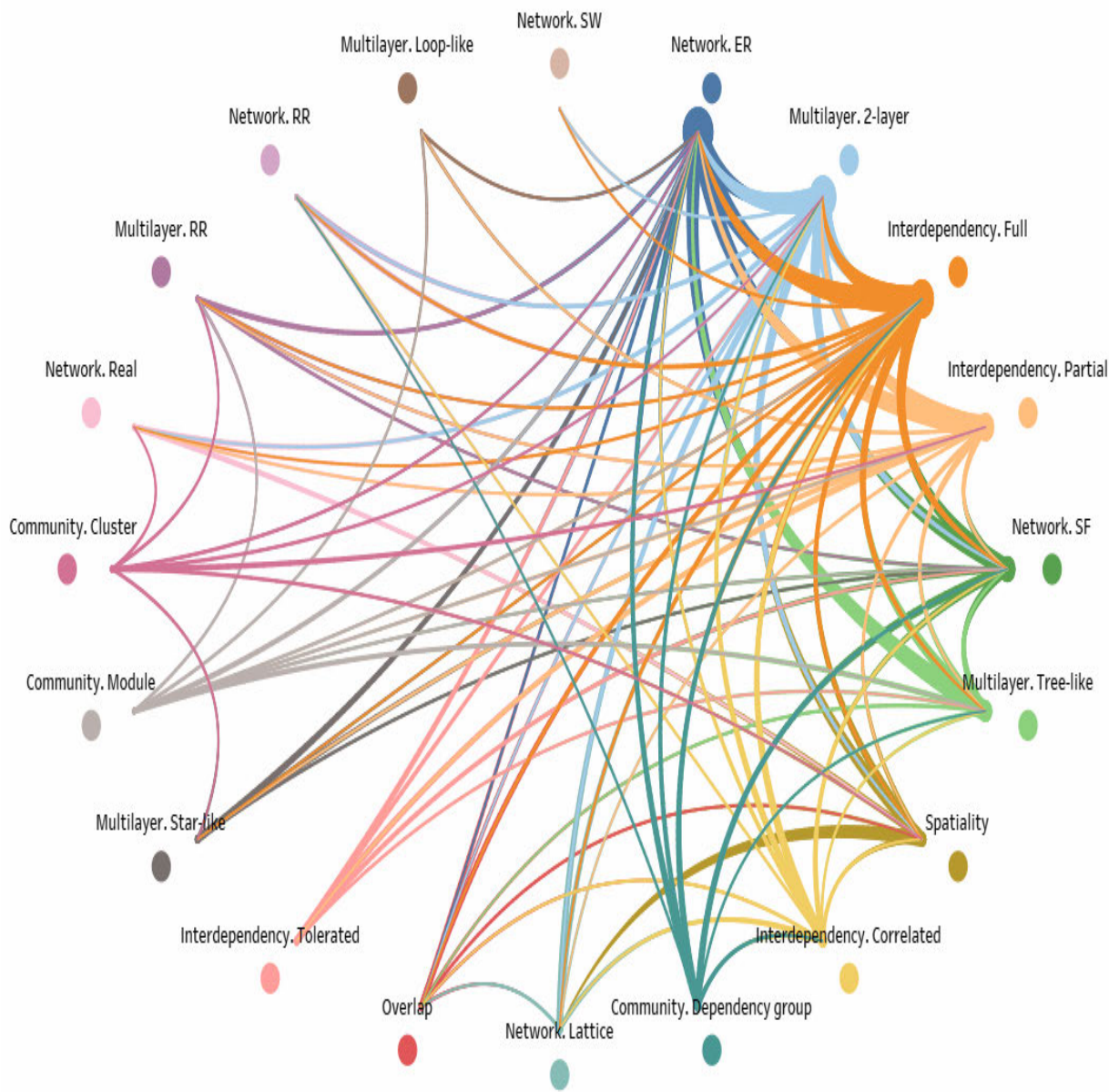

**FIGURE 14.** Schematic diagram to show the most and the least popular topological properties applied in the papers and how they were assembled to build the structure of multilayer models.

#### 2) THE EFFECT OF PERCOLATION MECHANISMS

While the type of attacks affects the robustness of multilayer networks, different mechanisms of failure percolation make no difference in how network properties affect the robustness of multilayer networks. For example, with more dependency links, the system is more vulnerable, regardless of different failure percolation mechanisms. However, comparing the deteriorative effect of the different percolation mechanisms, mentioned in the previous sections, needs more investigations in future studies.

### J. SIMPLIFICATION OF MODELS
#### 1) IN SYNTHETIC MODELS
Synthetic models of infrastructure systems were simplified by considering the same properties for layers e.g. same network, same number of nodes, same degree distribution, same number of communities, etc. while it is not the case in real-world infrastructure systems. In the studies in which researchers intended to evaluate the effect of more complicated properties such as overlaps, correlated interdependence, as well as spatial and temporal features, the multilayer models were limited to two layers.

From the 31 included papers, 27 of them only focused on the topological properties. However, only two of them considered functional and the two others worked on dynamical properties. Their models were also considered as a 2-layer model. Considering functional and dynamical properties in models is very crucial to capture the full characteristics of infrastructure systems. Flow of goods and services as well as specified load and capacity of components are essential parts of these systems [91]. Furthermore, the structure of infrastructure systems can evolve and change over time due to dynamic processes [92]. For example, in the most of studies, models of transportation systems used simplified static representations neglecting traffic flow, travel times, waiting times, and compatibility of schedules while using a dynamic representation that considers concepts like synchronizing, diffusion, and random walks obtained more accurate descriptions [93]–[95].

The overlap and spatiality were only studied for ER and lattice networks, respectively. No-feedback condition applied in all synthetic models while considering feedback condition is more realistic for infrastructure systems. All nodes were assumed to be connected by direct links while in many real cases, there is not a straight line between two components of infrastructure systems. Most of the multilayer models were built by ER networks while infrastructure systems are usually more complex and have non-homogeneous structures [17]. In total, among all models, 46% were 2-layer and tree-like models made by ER/SF networks with full/partial interdependency but without considering the presence of overlap, community, or spatiality (Figure 14).

#### 2) IN REALITY-BASED MODELS
Both papers with reality-based models applied 2-layer models [37], [38]. Only one of these two defined the dependency links with feedback conditions, where there were more than one dependency link connecting a node in the road network to some nodes in the sewer network [37]. Neither of these two models defined overlap in their real cases. According to these results, the inclusion of more properties results in models closer to real-world conditions. Likewise, considering more layers, unidirectional links, functional and dynamical properties, feedback conditions, and spatiality can make the models more realistic.

However, large, complex models have been criticized, mainly because they are computationally expensive and hard to implement, their results are usually hard to interpret, and their multipurpose approaches do not directly contribute to the development of theoretical aspects [96], [97]. On the other hand, simple models usually neglect important aspects of real-world systems such as heterogeneity and variability, and may focus too much on one of several properties [97]. Simple models are mostly generalized resulting in misleading modelers and may often be away from reality [98]. Nonetheless, while we need simple models to achieve and develop general insights and theories [97], [98], complex systems like infrastructures need complex solutions [99]. Specifically, some properties of complex systems like emergent behaviors as unexpected behaviors that stem from the multitude of interactions between different components cannot be analyzed by building simple models [100], [101].

Regarding the scope of this study, further works are needed to evaluate whether a model created by putting different network properties together would show unexpected behaviors or not. In other words, multilayer network models should contain different properties together to investigate if some properties boost or weaken the effect of other properties on the robustness and vulnerability of infrastructure systems. Since the results of the present study about the effect of different network properties on the robustness of infrastructure systems are mostly based on created simple models, it is needed to investigate whether these properties show the same behavior when they are put together in a complex model and when they are a focus of a simple model. In addition, the effect size of different properties on the robustness of infrastructure systems should be quantified to make more optimized decisions for improving infrastructure systems.

### V. CONCLUSION
Percolation theory is the most common approach to quantify the robustness of networked systems. However, the application of percolation theory to evaluate multilayer networks is still young and its use to analyze infrastructure systems is a major gap addressed in this review paper. All studies applying percolation theory to assess the robustness in infrastructure systems modeled as multilayer networks were collected. 19 network properties used in these studies were identified and some of these properties concerned vulnerability rather than robustness. Interdependency strength and communities were the most common network property whilst very few studies considered realistic attributes of infrastructure systems such as directed links and feedback conditions.

The review highlights that the properties led to approximately similar model outcomes, in terms of detecting improvement or deterioration in the robustness of multilayer infrastructure networks, although some exceptions were reported, like ineffectiveness of correlated interdependency in SF networks under dependency-based targeted attacks. Only two studies worked on real cases of infrastructure network data as most of the studies used simplified synthetic models of infrastructure systems. Each study assessed the effect of outages by focusing on one or two network properties and did not consider different network properties together.

In this regard, further studies are needed to analyze multiple properties in a single study to assess whether they boost or weaken the impact of each other. This would lead to a more comprehensive and decisive understanding about the effect of different properties under different conditions. Furthermore, the importance of each property on the robustness of infrastructure systems should be quantified in future studies to support the design and planning of robust infrastructure systems by arranging and prioritizing the most effective properties.

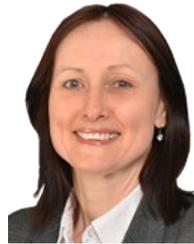

**LIZ VARGA** received the degree (Hons.) from The Open University, Milton Keynes, U.K., and the MBA and Ph.D. degrees from Cranfield University, Bedford, U.K. She has been the Chair of complex systems with University College London (UCL), London, U.K., since 2018, and was previously a Professor of complex infrastructure systems with Cranfield University, from 2015 to 2019. She leads the Infrastructure Systems Institute (UCL) and is the Section Head of infrastructure and cities with the Civil, Environmental and Geomatics Engineering Department, UCL. She is the Principal Investigator of the coordination node of U.K. Collaboratorium for Research in Infrastructure and Cities and is involved in several research projects and teaching commitments. She has published over 60 journal articles on infrastructure systems: energy, transport, water, waste, and telecommunications. She is a regular speaker, a reviewer, and an advisor on infrastructure matters, particularly sustainable innovation, resilience, and digitalization, and has advised various organizations and programmes, including the Royal Academy of Engineering, World Economic Forum, United Nations Office for Disaster Risk Reduction, and U.K.'s National Digital Twin Programme (NDTp). She is a Commissioner with the National Preparedness Commission. She is a fellow of the Chartered Institute of Building Engineers (FCABE) and the Higher Education Academy (HEA) and was awarded the Research Prize from Cranfield University, from 2014 to 2016.

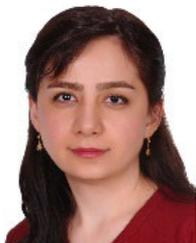

**ZAHRA MAHABADI** received the bachelor's degree in architecture from Shahrood University of Technology, Iran, in 2010, and the double master's degrees in urban and regional planning from the Art University of Isfahan, Iran, and in urban informatics from King's College London, U.K., in 2013 and 2018, respectively. She is currently pursuing the Ph.D. degree with University College London, U.K., with the focus of resilience of infrastructure through anticipatory and self-healing mechanisms. During her Ph.D. studies, she has been involved in a number of research projects in the areas of interdependent infrastructure systems and resilience principles.

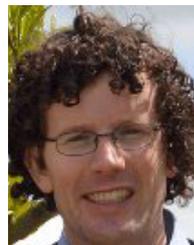

**TOM DOLAN** is a Senior Research Associate of UKCRIC Co-ordination Node with University College London (UCL). His research interests include complex infrastructure systems as enablers of societally beneficial outcomes. In particular, on the systemic interdependencies, emergent properties and dynamic context that underpin their normal operations and enable the realization of systemic outcomes that are net zero; sustainable and resilient to the disruptive impacts of global warming and other resilience challenges; enhance the quality of the local environment; and helps catalyse an urgent transformation to a net zero GHG emission economy. His research is systemic in scope and includes challenges of how we systemically govern, regulate, manage (including plan, design, procure, construct, operate, maintain, enhance, repurpose, measure, account for, value, and incentivise investment in) complex infrastructure systems for the safe emergence of these system characteristics.

● ● ●